\documentclass[aps,longbibliography,
  twocolumn,amsmath,
  amssymb,nofootinbib, superscriptaddress]{revtex4-1}

\usepackage{graphics}      
\usepackage{graphicx}      
\usepackage{longtable}     
\usepackage{url}           
\usepackage{bm}            
\usepackage{xcolor}
\usepackage{mathrsfs}
\usepackage[colorlinks,allcolors=blue]{hyperref}
\usepackage{blindtext}
\begin{document}
\title{Collective Oscillations of Majorana Neutrinos in Strong Magnetic Fields and  Self-induced Flavor Equilibrium}

\author{Sajad Abbar}
\affiliation{Max-Planck-Institut f\"ur Physik (Werner-Heisenberg-Institut),\\ F\"ohringer Ring 6, 80805 M\"unchen, Germany}
\affiliation{Astro-Particule et Cosmologie (APC), CNRS UMR 7164, Universit\'e Denis Diderot, 75205 Paris Cedex 13, France }


\begin{abstract}
We study collective oscillations of Majorana neutrinos in some of the most extreme astrophysical sites such as neutron star merger remnants and magneto-rotational core-collapse supernovae which include dense neutrino media in the presence of strong magnetic fields. We show that
 neutrinos can
 reach  flavor equilibrium
  if  neutrino transition magnetic moment $\mu_\nu$
   is strong enough,
 namely when  $\mu_\nu/\mu_{\rm{B}} \gtrsim 10^{-14}-10^{-15}$
  with $\mu_{\rm{B}}$ being the Bohr magneton. 
This  sort of flavor equilibrium,  which is not necessarily flavor equipartition,
can occur on  (short) scales determined by the strength of the magnetic
 term in the Hamiltonian.
Our findings can have interesting implications for the physics of
 such violent astrophysical environments. 

\end{abstract}

\maketitle


\section{Introduction}
Neutrinos can play a vital role in the physics of
the most violent astrophysical phenomena such as
 neutron star mergers (NSM) and
 core-collapse  supernovae (CCSNe)~\cite{Colgate:1966ax, Bethe:1984ux, 
 Janka:2012wk, Burrows:2012ew}.
 Due to their weak interactions, 
they can act as the major channel of  energy transport.
Moreover,  they can be crucial to
 heavy elements nucleosynthesis since they can modify
the neutron to proton ratio through the weak reactions
$\bar\nu_e + p \rightleftharpoons n + e^+$ and
$\nu_e + n \rightleftharpoons p + e^-$~\cite{Qian:1996xt}.

Neutrinos can  experience flavor conversions
which  can change their energy spectra.
This, in principle, can  change 
 their  interaction rates 
and consequetly influence their effects on the dynamics and
nucleosynthesis in the extreme astrophysical environments. 
In addition, on the observational side, any flavor conversions  can
modify the neutrino signal which may be observed from
these events on the earth.

  Neutrinos can experience collective flavor
oscillations in NSM remnants and CCSNe due to 
their coherent forward scatterings by
the high  density background neutrino gas.   
The presence of neutrino-neutrino interaction makes the 
problem of neutrino evolution in a dense neutrino medium very
demanding and remarkably different from the one in vacuum and matter. 
It, indeed, makes  this problem a nonlinear one with  strong 
 coupling among different neutrino momenta~\cite{Pastor:2002we,duan:2006an, duan:2006jv, duan:2010bg,
  Chakraborty:2016yeg}. 

The first studies on this problem where carried out in  maximally
symmetric models.
 For example, to study collective neutrino
oscillations in the supernova context, a stationary spherically symmetric  
SN model, i.e. the so-called neutrino bulb model
was used~\cite{duan:2006an}. 
The most important feature of the results obtained in 
the bulb  model is the presence of the spectral swapping phenomenon 
in which   $\nu_e$ ($\bar{\nu}_e$)  exchanges its spectra with 
$\nu_x$ ($\bar{\nu}_x$)  for a certain range of neutrino 
energies~\cite{duan:2006jv, duan:2007sh, dasgupta:2009mg, duan:2010bg,
 Galais:2011gh, Duan:2007bt, Galais:2011gh, Duan:2015cqa}.
This phenomenon is a direct consequence of 
collective neutrino oscillations.  

 However, regarding the evolution of neutrinos in NSM
remnant accretion disks,
 the geometry is much more complicated and  a self-consistent
one-dimensional model is unavailable. 
The first studies were done in the so-called
single-angle approximation scenario~\cite{Malkus:2012ts} in which it is assumed 
that all  neutrinos emitted from the neutrino emitting accretion disk
 experience similar flavor evolution.
The salient characteristic of the results obtained in 
these calculations is the occurrence of matter-neutrino 
resonance (MNR)~\cite{Malkus:2014iqa, Malkus:2015mda, Vlasenko:2018irq,
Shalgar:2017pzd, Tian:2017xbr, Frensel:2016fge, Zhu:2016mwa, Wu:2015fga,
Chatelain:2016xva, Chatelain:2017yxx}.
This phenomenon results from 
 the cancellation between the neutrino-neutrino interaction and matter potentials
 which can happen since in the NSM environment, 
 $\bar{\nu}_e$ can be more abundant than ${\nu}_e$,
  i.e. $n_{\bar{\nu}_e}/n_{{\nu}_e} > 1$. 
 The MNR phenomenon is currently thought to be absent 
  in the supernova environment where normally $n_{\bar{\nu}_e}/n_{{\nu}_e} < 1$
 and as a result,
 the neutrino and matter potentials have similar signs\footnote{
 Note that there can exist SN zones
 inside the proto-neutron star for which  
 $n_{\bar{\nu}_e}/n_{{\nu}_e} > 1$~\cite{DelfanAzari:2019tez, Abbar:2019zoq, Glas:2019ijo}.
 Despite this, the matter density is much larger than the neutrino 
 number densities at these zones and the cancelation
 between the neutrino-neutrino interaction and matter potentials seems 
 to be inacssisible.}.
  
 Nevertheless, it was then realised that such oversimplified
 maximally symmetric models are 
 not appropriate to study neutrino flavor evolution 
 in dense neutrino media. On the one hand, the spatial and time
 symmetries in the neutrino gas can be broken spontaneously in 
 the presence of collective neutrino 
 oscillations~\cite{raffelt:2013rqa, duan:2013kba, duan:2014gfa,
 abbar:2015mca, Abbar:2015fwa, chakraborty:2015tfa, Chakraborty:2016yeg,
 Dasgupta:2015iia, Mirizzi:2015fva,Martin:2019gxb, Martin:2019kgi}. 
 This can
 allow for neutrino flavor conversions at very large matter/neutrino
 densities. On the other hand, it has been shown that 
 neutrinos can experience the so-called fast flavor conversion 
 modes in dense neutrino media probably provided that $\nu_e$ and $\bar\nu_e$
 angular distributions cross each other~\cite{Sawyer:2005jk, Sawyer:2015dsa,
 Chakraborty:2016lct, Izaguirre:2016gsx,  Wu:2017qpc,
  Capozzi:2017gqd, Richers:2019grc,  
  Dasgupta:2016dbv, Abbar:2017pkh, Abbar:2018beu, Capozzi:2018clo,
 Martin:2019gxb, Capozzi:2019lso, Doring:2019axc, Chakraborty:2019wxe, Johns:2019izj,
 Shalgar:2019qwg, Cherry:2019vkv}. 
 Such fast conversion modes
 can occur on scales $\sim G_{\mathrm{F}}^{-1} n_\nu ^{-1}$
which can be as short as a few
 cm's in the aforementioned extreme astrophysical environments.  
 This must be compared with  
   slow modes expected to
 occur on scales $\sim  \mathcal O  (1)$ km (for a  10 MeV  neutrino) 
 determined
by  the neutrino vacuum frequency $\omega = \Delta m_{\mathrm{atm}}^2/2E $.

In addition, neutrinos are expected to have tiny but nonzero magnetic 
moments (see, e.g., ~\cite{Giunti:2008ve, Broggini:2012df, Studenikin:2016ykv} for a review)
which can influence their flavor evolution  in the presence of  magnetic fields.  
In particular, the presence of ultra-strong magnetic fields ($B \gtrsim 10^{15}$ 
Gauss~\cite{Mosta:2015ucs})  in NSM
and magneto-rotational CCSNe (in which rapid rotation and large magnetic
fields are thought to play an important role) makes them ideal settings 
for studying the impact of  
the coupling between neutrinos  and
 magnetic field (photon) on collective neutrino oscillations.
 While such a coupling leads to
  \emph{active-sterile} neutrino oscillations
in the case of Dirac neutrinos,  
 it results in \emph{neutrino-antineutrino}  oscillations
for Majorana neutrinos. 

In the minimally-extended Standard Model (MESM),
 the diagonal  magnetic
moment of  Dirac neutrinos can be written as~\cite{Fujikawa:1980yx}
\begin{equation}
\mu^{ii,D}_\nu = \frac{3eG_{\rm{F}} m_\nu} {8\sqrt2 \pi^2} 
\simeq 3.2 \times 10^{-19}\ (\frac{m_\nu}{1 \rm{eV}})\ \mu_{\rm{B}},
\end{equation}
where $m_\nu$ is the neutrino mass and $\mu_{\rm{B}} = 5.788\times 10^{-9}$
eV Gauss$^{-1}$ is the Bohr magneton. The transition magnetic
moment is smaller than 
the diagonal one by approximately four orders of magnitude.
As for Majorana neutrinos, while the diagonal magnetic moment is 
dictated to be zero,  the transition magnetic moment is similar to 
the transition magnetic moment of Dirac neutrinos.

Although MESM predicts $\mu_\nu \lesssim 10^{-19} \mu_{\rm{B}}$, 
some of the theories beyond SM predict (or at least 
can explain)  much larger values for 
$\mu_\nu$\footnote{There can indeed exist some difficulties 
here~\cite{Lindner:2017uvt, Bell:2005kz, Davidson:2005cs}.
In particular, since the neutrino magnetic moment can depend linearly on 
the neutrino mass, any attempt to increase the neutrino magnetic
moment  leads to an increase in the neutrino mass as well.
Thus, to be consistent with current constraints on the neutrino
mass and at the same time having large values for $\mu_\nu$,
 one may need a sort of fine-tuning.
 Though for Dirac neutrinos
this necessary fine-tuning leads to theoretical difficulties to produce $\mu_\nu \gtrsim 10^{-15}  \mu_{\rm{B}}$,
it is almost harmless to the case of Majorana neutrinos since
it only becomes problematic when $\mu_\nu \gtrsim 10^{-9}  \mu_{\rm{B}}$
which is already excluded by experiments. }. 
In fact, 
the current
experiments can only provide an upper bound 
on $\mu_\nu$ (see, e.g., Refs.~\cite{Canas:2015yoa, Beda:2012zz}),
\begin{equation}
\mu_\nu \lesssim 3 \times 10^{-11} \mu_{\rm{B}},
\end{equation}
which is many orders of magnitude larger than the value suggested by MESM.
This constraint is valid for both Dirac and Majorana
neutrinos and also
diagonal and transition magnetic moments.

The coupling between  neutrinos 
 and  magnetic field
can provide new channels for changing
neutrino lepton number and can possibly lead to new physics, if it is
strong enough.
A number of  papers have studied this phenomenon in astrophysical
environment~\cite{Kolb:1981mc, Schechter:1981hw,
Lim:1987tk, Balantekin:1990jg,  Akhmedov:1992ea, Akhmedov:1997qb,
Ando:2002sk, Lychkovskiy:2009pm,
Balantekin:2007xq, Kuznetsov:2009we, Akhmedov:2003fu, 
deGouvea:2012hg, deGouvea:2013zp, Dvornikov:2011dv, 
Dobrynina:2016rwy, 
  Kurashvili:2017zab}.
In particular, in Refs.~\cite{deGouvea:2012hg, deGouvea:2013zp} 
 the authors reported  that  collective  oscillations of Majorana neutrinos can
be nontrivially affected by the magnetic term  for 
level-of-SM or even smaller  $\mu_\nu$'s. 

In this paper,
 we study collective oscillations of Majorana neutrinos in the
presence of strong magnetic fields with $B \gtrsim 10^{15}$ Gauss,
thought to be present in 
NSM remnants and magneto-rotational  CCSNe.
To achieve this goal, we use a schematic multi-angle one-dimensional model
for the neutrino gas
in the two-flavor (Sec.~\ref{sec:2f}) and three-flavor (Sec.~\ref{sec:3f}) scenarios.
 We show that if the neutrino 
magnetic moment is large enough,  
the neutrino gas can reach a sort of flavor equilibrium (which is not
necessarily equipartition) on scales determined by the magnetic term. 

\section{Collective oscillations of Majorana neutrinos in the presence of magnetic fields} 

To study the evolution of Majorana neutrinos in the presence of strong magnetic fields,
we consider a  single-energy, multi-angle neutrino gas in both two and three-flavor
scenarios in which neutrinos 
are emitted with emission angles in the range 
$[-\vartheta_{\rm{max}}, \vartheta_{\rm{max}}]$. This model is similar to the
one used in Ref.~\cite{Abbar:2018beu}.

At each space-time point $(t,\bold{r})$, the flavor state of a neutrino 
 traveling in direction $\vartheta$ 
  can be specified by its density matrix $\rho_{\vartheta}(t,\bold{r})$.
   The evolution of $\rho_{\vartheta}(t,\bold{r})$
in the absence of collisions is governed  by the 
Liouville-von Neumann equation of motion~\cite{Sigl:1992fn,Strack:2005ux,Cardall:2007zw,Volpe:2013jgr, Vlasenko:2013fja}
\begin{equation}
i    \rm{D}_t  \rho_{ \vartheta} = [\rm{H}_{\vartheta},\rho_{ \vartheta}],
\label{eq:EOM}
\end{equation}
where $\rm{D}_t = \partial_t + \bold{v}\cdot{\bold{\nabla}}$ and
$\rm{H}_{\vartheta} = \rm{H_{vac}} + \rm{H_{mat}} + \rm{H}_{\nu \nu, \vartheta}$ 
is the total Hamiltonian, with $\rm{H_{vac}}$, $\rm{H_{mat}}$ 
and $\rm{H}_{\nu \nu, \vartheta}$ being the contributions 
from vacuum, matter and neutrino-neutrino interaction potentials, respectively.
Here, the contribution from the coupling between  neutrinos 
 and  magnetic field is included in the vacuum term.

In our study, the evolution of neutrinos is considered in 
two models, namely a stationary one-dimensional model and 
a time-dependent homogenous neutrino gas. In the one-dimensional
model $\rm{D}_t = \cos\vartheta d_r$ while one has $\rm{D}_t =  d_t$ in the
time-dependent homogenous gas.
As will be seen in what follows,
the occurrence and  nature of the equilibrium  does not depend on the employed model
since the outcome is purely determined by the presence of the 
strong magnetic coupling term. Nevertheless, the amplitude of the oscillations around the
equilibrium can be smaller in the stationary one-dimensional model.
We also assume that 
the physical quantities such as the matter/neutrino densities
and magnetic field are constant during the propagation
of neutrinos. This is justified by noting that the scales
associated with neutrino oscillations in this problem (induced by strong
magnetic coupling) are much shorter
than the relevant  scales of the astrophysical problems
of interest.

\subsection{Two-flavor scenario}\label{sec:2f}

To demonstrate the idea  and to show how the 
presence of strong coupling between  neutrinos 
 and  magnetic field can influence their oscillations
in a dense neutrino medium, we first start with the case of
two-flavor scenario.
We follow the formalism developed in Ref.~\cite{deGouvea:2012hg, deGouvea:2013zp}
  and we take $\rho$ to be a $4\times4$ matrix
   which includes the flavor content  
    of  neutrinos and antineutrinos
  \begin{equation}
\rho = 
\left[ {\begin{array}{cccc}
\rho_{\nu_e\nu_e}  &  \rho_{\nu_e\nu_x}  & \rho_{\nu_e \bar{\nu}_e} & \rho_{\nu_e \bar{\nu}_x}   \\
\rho_{\nu_x\nu_e}  &  \rho_{\nu_x\nu_x}  & \rho_{\nu_x \bar{\nu}_e} & \rho_{\nu_x \bar{\nu}_x}   \\
\rho_{\bar{\nu}_e\nu_e}  &  \rho_{\bar{\nu}_e \nu_x}  & \rho_{\bar{\nu}_e \bar{\nu}_e} & \rho_{\bar{\nu}_e \bar{\nu}_x}   \\
\rho_{\bar{\nu}_x \nu_e}  &  \rho_{\bar{\nu}_x \nu_x}  & \rho_{\bar{\nu}_x \bar{\nu}_e} & \rho_{\bar{\nu}_x \bar{\nu}_x}   \\
\end{array} } \right],
\end{equation}
where the diagonal 
terms are basically the occupation numbers of the
corresponding neutrino flavors and off-diagonal terms carry information 
on neutrino flavor mixing. 
This matrix  has clearly the form
\begin{equation}
\rho = 
\left[ {\begin{array}{cc}
\rho_{\nu}  & X   \\
X^\dagger  &  \rho_{\bar\nu}     \\
\end{array} } \right],
\end{equation}
with 
\begin{equation}
X = \left[ {\begin{array}{cc}
 \rho_{\nu_e \bar{\nu}_e} & \rho_{\nu_e \bar{\nu}_x}   \\
\rho_{\nu_x \bar{\nu}_e} & \rho_{\nu_x \bar{\nu}_x}  \\
\end{array} } \right],
\end{equation}
and $\rho_{\nu}$ and  $\rho_{\bar\nu} $ being  the usual $2\times2$ flavor
matrices having information on the flavor content of neutrinos and
antineutrinos, respectively.
It is very convenient to
follow this formalism here
since for nonzero Majorana neutrino magnetic moment, 
 neutrinos and antineutrinos are coupled in the presence of
magnetic field and   there is a nonzero
$\nu-\bar\nu$ transition amplitide.

 Within this formalism, the vacuum and matter potentials can be written as
\begin{align}
\rm{H_{vac}} &= 
\left[ {\begin{array}{cccc}
-\omega \cos2\theta_{\textrm{v}}  &   \omega \sin2\theta_{\textrm{v}}  & 0 & \Omega  \\
 \quad \omega \sin2\theta_{\textrm{v}}   & \omega \cos2\theta_{\textrm{v}} & - \Omega & 0 \\
  0 & - \Omega &  -\omega \cos2\theta_{\textrm{v}}  &   \omega \sin2\theta_{\textrm{v}} \\
  \Omega & 0 & \quad \omega \sin2\theta_{\textrm{v}}   & \omega \cos2\theta_{\textrm{v}} \\
\end{array} } \right], \\
 \rm{H_{mat}}  &=
\left[ {\begin{array}{cccc}
( \lambda_e - \lambda_n/2)   &   0  & 0 & 0\\
0  & -  \lambda_n/2  & 0 & 0 \\
0 & 0 & -( \lambda_e - \lambda_n/2) & 0\\
0 & 0 & 0 & \lambda_n/2 \\
\end{array} } \right], 
\end{align}
where $\lambda_{\rm{e(n)}} = \sqrt2 G_{\mathrm{F}} n_{e(n)}$, with $n_e$ ($n_n$)  
being the electron (neutron) number density and 
$\theta_{\textrm{v}}$  and $\omega = \Delta m_{\mathrm{atm}}^2/2E$
are the neutrino vacuum mixing angle and the vacuum 
frequency ($\Delta m_{\mathrm{atm}}^2 > 0$ ($ < 0$) for 
 the normal (inverted) mass hierarchy) for a 
neutrino with energy $E$. In our calculations, we set $\theta_{\textrm{v}}=0.1$ and $\omega=1$
though the results do not qualitatively depend on the choice of these parameters.
Note that the vacuum term has the new contribution $\Omega = \mu_\nu B_{\rm{T}}$ 
 from the coupling of Majorana neutrino 
  with the component of 
magnetic field transverse to the neutrino momentum, $B_{\rm{T}}$. 
Furthermore, unlike the case of collective neutrino oscillations in the absence
of magnetic field, the neutral current contribution from neutrons to
the matter potential can not be ignored since
 it has different signs for neutrinos and antineutrinos
 and can not be removed as a common phase when 
 these two are coupled.
 
In addition, the neutrino-neutrino interaction potential, $\rm{H}_{\nu \nu, \vartheta}$, 
is\footnote{Here and in Eq. (8), we assume having a large $\mu_\nu$ does not
modify neutrino weak interactions (similar to, e.g., 
Refs.~\cite{Akhmedov:1992ea, Akhmedov:1997qb}).}
\begin{equation}
\begin{split}
 \mathrm{H}_{\nu \nu,\vartheta} &=\sqrt2 G_{\rm{F}} \int_{-\vartheta_{max}}^{\vartheta_{max}} \mathrm{d}\vartheta' \quad  \big(1- \cos(\vartheta - \vartheta')\big)\\
& \times  [G^{\dagger} ( \rho_{\vartheta'} - \rho^{c*}_{\vartheta'} ) G
+ \frac{1}{2} G^{\dagger} \rm{tr}\big( ( \rho_{\vartheta'} - \rho^{c*}_{\vartheta'}  )G \big)],
 \end{split}\label{eq:Hvv}
 \end{equation}
where 
\begin{equation}
G = 
\left[ {\begin{array}{cccc}
+1 &  0  & 0 & 0   \\
0 &  +1 & 0 & 0   \\
 0&0  &-1 &0  \\
  0&    0 & 0 &  -1 \\
\end{array} } \right],
\end{equation}
and $\rho^c$ is defined as
\begin{equation}
\rho^c = 
\left[ {\begin{array}{cc}
\rho_{\bar\nu}  & X^*   \\
X^T  &  \rho_{\nu}     \\
\end{array} } \right].
\end{equation}
Note that the definition of $\rho^c$ is somewhat different from the one in 
Refs.~\cite{deGouvea:2012hg, deGouvea:2013zp} so that there is no contribution to $\nu - \bar{\nu}$
transition from  the neutrino-neutrino interaction 
term~\cite{ Dvornikov:2011dv, Cirigliano:2014aoa} (see also \cite{Vaananen:2013qja, Volpe:2015rla, Vlasenko:2013fja}).
The last term in Eq.~(\ref{eq:Hvv}) refers to a phase factor which
has different signs for neutrinos and  antineutrinos and therefore, 
can not be removed here.
 One can then recover the usual equations of motion of 
traditional collective oscillations if $B=0$.

\subsubsection{Results}
 In our simulations, we took
$\vartheta_{\rm{max}} = \pi/3$ and  a fixed magnetic field  
 with $B_{\rm{T}} = 5 \times10^{15}$ Guass. Such strong magnetic
 fields may not exist on very large scales in the astrophysical 
 problems of interest. However, 
 the scales  associated with neutrino oscillations  for strong 
 $\Omega$'s  are much shorter than other relevant scales in the problem
and therefore, we here  intend to consider the local effects of large   $\Omega$'s
 rather than the global ones. 
 Thus, the physical quantities are assumed to be constant.
 Note also that
since  Hamiltonian is only sensitive to $B$ via  $ \mu_\nu B_{\rm{T}}$, for
 smaller/larger magnetic fields one can just rescale 
 $\mu_\nu$.
We also set $n_{\nu_\mu} = n_{\bar\nu_\mu} = n_{\nu_\tau} = n_{\bar\nu_\tau} = 0.4\ n_{\nu_e}$
in our calculations.

\begin{figure*} [tbh!]
 \centering
\begin{center}
\includegraphics*[width=.97\textwidth,trim=0 25 10 10,clip]{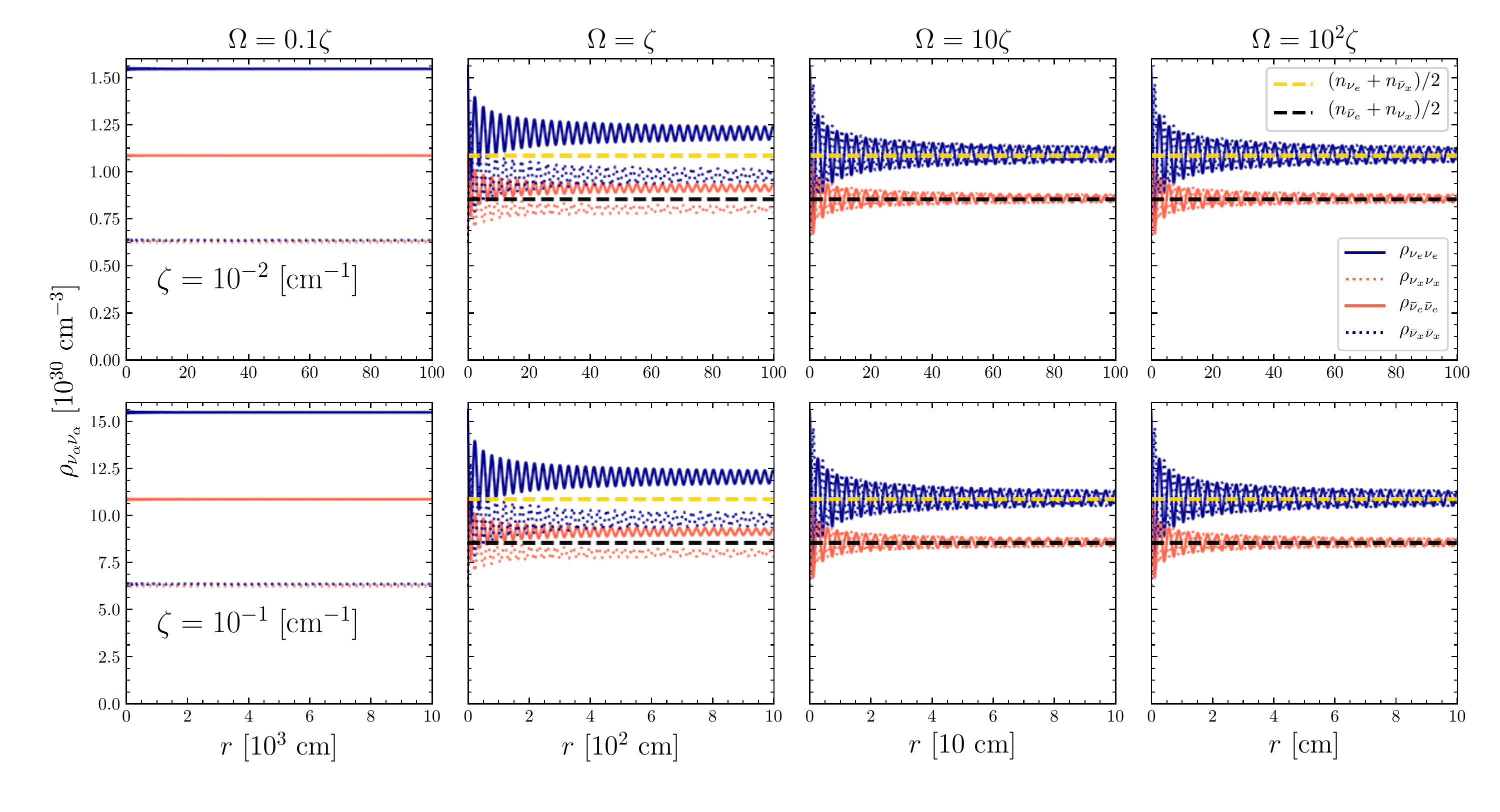}
\end{center}
\caption{   Angle-averaged  survival
probabilities of neutrinos and antineutrinos 
 for $n_{\bar\nu_e}/n_{\nu_e}=0.7$. The evolution of neutrinos
 is studied in the homogenous time-dependent model and as mentioned 
 in the text, $\mu_\nu$ can be found from $\mu_\nu = 6.8\times10^{-15}\mu_{\rm{B}} \  
 \big({\Omega}/{\zeta}\big) \big({\zeta}/{10^{-2}\ \rm{cm}^{-1}}\big)$ in each
 panel. Note that the neutrino oscillations scale is $\sim 1/\Omega$ for
 strong $\Omega$'s. } 
\label{fig:alpha0.7}
\end{figure*}

\begin{figure*} [tbh!]
 \centering
\begin{center}
\includegraphics*[width=.97\textwidth,trim=10 25 10 10,clip]{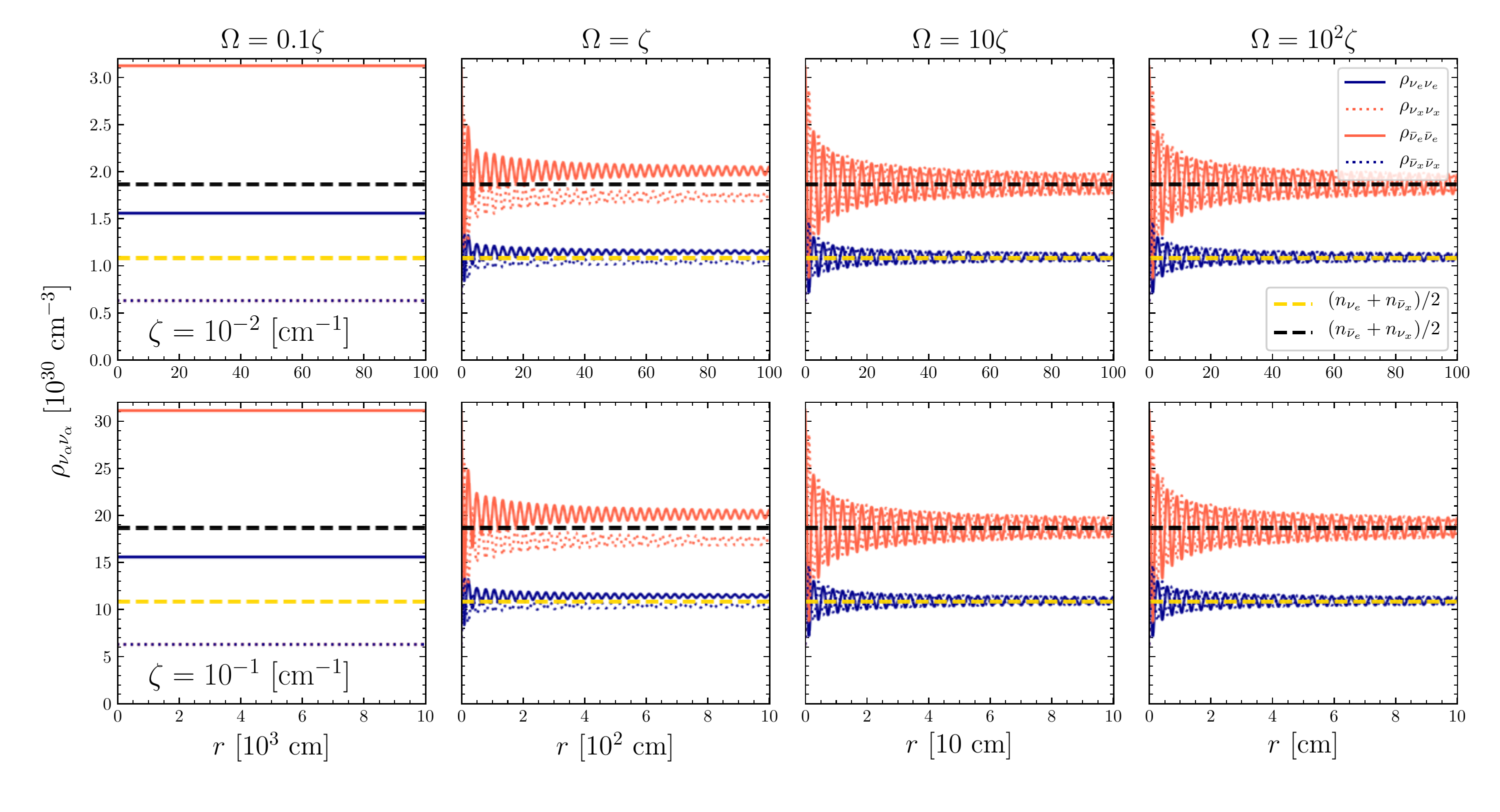}
\end{center}
\caption{  The same information as in Fig.~\ref{fig:alpha0.7} for 
$n_{\bar\nu_e}/n_{\nu_e}=2$.} 
\label{fig:alpha2}
\end{figure*}

The angle-averaged neutrino survival probabilities
of neutrinos and antineutrinos are shown in
Figs.~\ref{fig:alpha0.7} and \ref{fig:alpha2}. We considered 
 two 
 cases with $n_{\bar{\nu}_e}/n_{\nu_e} = 0.7 $
and $2$, for a number of  $\Omega$'s and two neutrino 
number densities specified by
\begin{equation}
\zeta = \sqrt2 G_{\mathrm{F}} n_{\nu_e}.
\end{equation}
For each panel, the corresponding neutrino magnetic moment is
\begin{equation}
\mu_\nu = 6.8\times10^{-15}\mu_{\rm{B}} \  \big(\frac{\Omega}{\zeta}\big) \big(\frac{\zeta}{10^{-2}\ \rm{cm}^{-1}}\big).
\end{equation}
We have
confirmed that our results do not qualitatively depend on the choice
of $n_{\bar{\nu}_e}/ n_{\nu_e} $ and $n_{\nu_x}/ n_{\nu_e} $, as
well as electron and neutron densities (as long as $\Omega$
is the dominant term) and the mass term in the Hamiltonian.

\begin{figure*} [tbh!]
 \centering
\begin{center}
\includegraphics*[width=.97\textwidth,trim=0 10 10 5,clip]{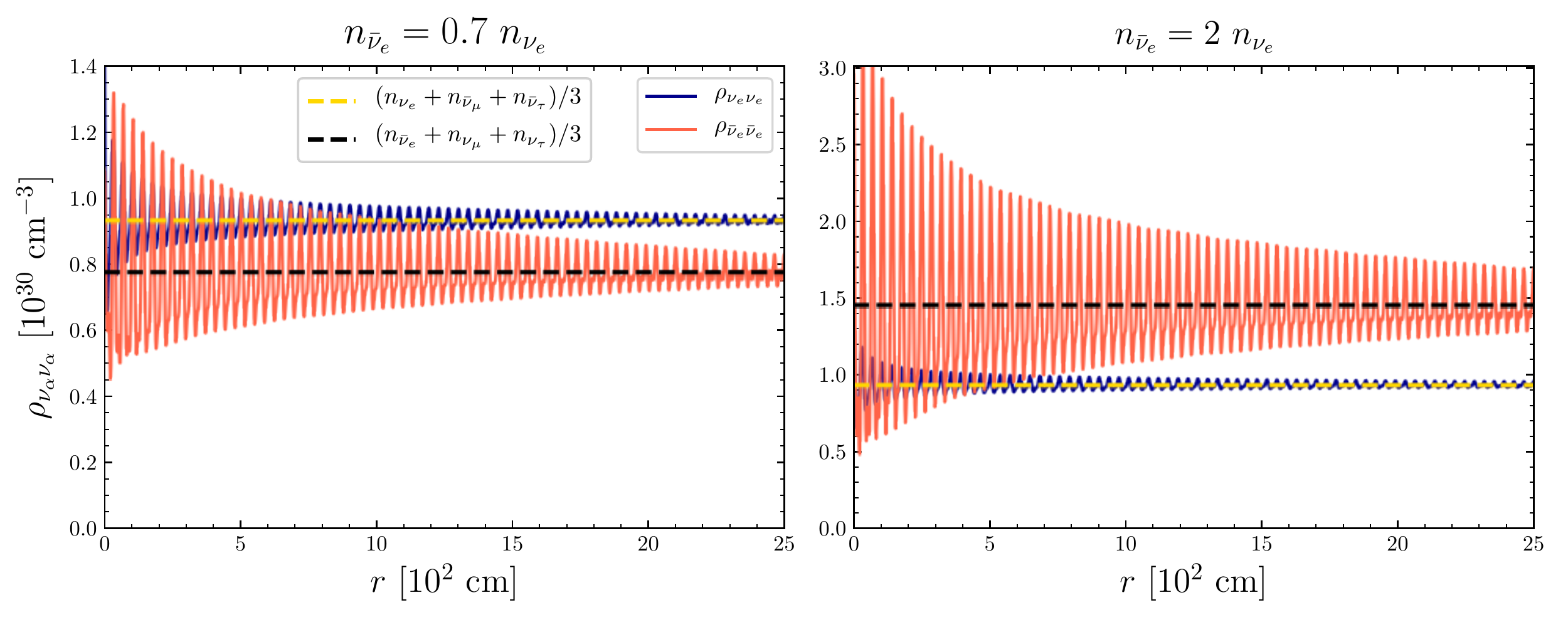}
\end{center}
\caption{  Angle-averaged  survival
probabilities of neutrinos and antineutrinos in the three-flavor
calculation in the stationary one-dimensional model
 with $\zeta=10^3$, $\Omega=10\zeta$ and two 
different values of $n_{\bar\nu_e}/n_{\nu_e}$. A similar flavor equilibrium
can occur among the coupled neutrino/antineutrino species.
$\Omega_{\alpha\beta}$'s are assumed to be equal here.} 
\label{fig:3f}
\end{figure*}

As can be clearly seen in Figs.~\ref{fig:alpha0.7} and \ref{fig:alpha2}, 
the magnetic term does not noticeably  modify neutrino flavor evolution for
small $\Omega$'s ($\Omega \lesssim 0.5 \zeta$). But
if  $\Omega$ term is comparable to the rest of the Hamiltonian
(here dominated by $\rm{H}_{\nu \nu}$),
 the neutrino gas experiences an  interesting sort of flavor
 equilibrium in which 
$\nu_e$ ($\bar{\nu}_e$) reaches an approximate equalisation with  $\bar{\nu}_x$ ($\nu_x$)
so that  
\begin{equation}
\begin{split}
\rho_{\nu_e \nu_e} \simeq \rho_{\bar{\nu}_x \bar{\nu}_x}  \simeq \frac{ n_{\nu_e}  + n_{\bar{\nu}_x} }{2}\\
\rho_{\bar{\nu}_e \bar{\nu}_e} \simeq \rho_{\nu_x \nu_x} \simeq \frac{ n_{\bar{\nu}_e} + n_{\nu_x} }{2}\\
\end{split}
\end{equation}
with some small amplitude oscillations around the equilibrium value
which can become smaller
 for  $\Omega >  \sqrt2 G_{\rm{F}} n_{\nu_e} $.
 The from of flavor equilibrium is indeed independent  
 of the ratio $n_{\bar{\nu}_e}/n_{\nu_e}$.

The  special form of the equilibrium  arises from 
 the specific structure of the vacuum Hamiltonian
which couples $\nu_e \leftrightarrow \bar{\nu}_x$ and 
$\bar{\nu}_e \leftrightarrow   \nu_x $. For strong 
$\Omega$'s, the vacuum term dominates the evolution of 
neutrinos meaning that the neutrino oscillations is caused 
and triggered by the magnetic term
which implies that the neutrino oscillation frequency is here determined
by $\Omega$.
This, combined with the decoherence induced  by
the neutrino-neutrino interaction term can then lead to the flavor equilibrium.
Note that here the flavor conversion  does not arise from 
 cancellation between the diagonal
terms in the Hamiltonian as in Refs.~\cite{Lim:1987tk, Akhmedov:1997qb}, where resonant 
conversion is responsible for neutrino flavor
oscillations.

\subsection{Three-flavor scenario}\label{sec:3f}

Three-flavor oscillations of a dense neutrino gas in the presence of strong
coupling between  neutrinos  
 and  magnetic field can be studied as a straightforward 
generalisation of the  two-flavor case. 

The $6\times6$  neutrino density matrix 
 \begin{equation}
\rho = 
\left[ {\begin{array}{cc}
\rho_{\nu}  & X   \\
X^\dagger  &  \rho_{\bar\nu}     \\
\end{array} } \right],
\end{equation}
 includes the flavor contents  
 of both neutrinos and antineutrinos of all three flavors
with 
\begin{equation}
X = \left[ {\begin{array}{ccc}
 \rho_{\nu_e \bar{\nu}_e} & \rho_{\nu_e \bar{\nu}_\mu}  &  \rho_{\nu_e \bar{\nu}_\tau}   \\
\rho_{\nu_\mu \bar{\nu}_e} & \rho_{\nu_\mu  \bar{\nu}_\mu}  &  \rho_{\nu_\mu  \bar{\nu}_\tau}   \\
\rho_{\nu_\tau \bar{\nu}_e} & \rho_{\nu_\tau \bar{\nu}_\mu}  &  \rho_{\nu_\tau \bar{\nu}_\tau}   \\
\end{array} } \right],
\end{equation}
and $\rho_{\nu}$ and  $\rho_{\bar\nu} $ are  the usual $3\times3$ flavor
matrices of neutrinos and
antineutrinos, respectively, defined as
 \begin{equation}
\rho = 
\left[ {\begin{array}{ccc}
\rho_{\nu_e\nu_e}  &  \rho_{\nu_e\nu_\mu}  & \rho_{\nu_e \nu_\tau}  \\
\rho_{\nu_\mu \nu_e}  &  \rho_{\nu_\mu \nu_\mu}  & \rho_{\nu_\mu \nu_\tau}    \\
\rho_{\nu_\tau \nu_e}  &  \rho_{\nu_\tau \nu_\mu}  & \rho_{\nu_\tau \nu_\tau}   \\
\end{array} } \right],
\end{equation}
and similarly for antineutrinos.

 In addition, the vacuum Hamiltonian can be written as
 \begin{equation}
\rm{H_{vac}} = 
\left[ {\begin{array}{cc}
{\tilde{H}_{\rm{vac}}} &  \tilde{H}_{\rm{B}}  \\
-\tilde{H}_{\rm{B}}  &  \tilde{H}_{\rm{vac}}^*   \\
\end{array} } \right],
\end{equation}
where ${\tilde{H}_{\rm{vac}}}$ is the usual $3\times3$  three-flavor vacuum 
Hamiltonian described by two mass-squared differences  
$\Delta m_{\mathrm{12}}^2$ and $\Delta m_{\mathrm{13}}^2$,
three mixing angles $\theta_{12}$, $\theta_{13}$ and $\theta_{23}$,
and  one  \textit{CP}-violating phase 
$\delta$\footnote{Although we set $\delta=0$ in our calculations,
we have confirmed that the results do not qualitatively depend on the choice 
of $\delta$.}, for which 
the values were taken from  Particle Data Group~\cite{Tanabashi:2018oca}. 
Also, 
\begin{equation}
\tilde{H}_{\rm{B}} = 
\left[ {\begin{array}{ccc}
0 &  \Omega_{e\mu} & \Omega_{e\tau}  \\
 -\Omega_{e\mu} &   0 &\Omega_{\mu\tau}  \\
 -\Omega_{e\tau} &   -\Omega_{\mu\tau} &0  \\
\end{array} } \right],
\end{equation}
describes the contribution from the magnetic term where
$\Omega_{\alpha\beta} = \mu_{\alpha\beta} B_{\rm{T}}$ are
assumed  to be real quantities. Moreover, in the neutrino-neutrino interaction
term, Eq.~(\ref{eq:Hvv}), $G$ and $\rho^c$ are  straightforward   $6\times6$
generalisations of the corresponding $4\times4$ ones,
\begin{equation}
G = 
\left[ {\begin{array}{cccccc}
+1 &  0  & 0 & 0 & 0 &0   \\
0 &  +1 & 0 & 0  & 0 &0  \\
 0&0  & +1 &0  & 0 &0  \\
  0&    0 & 0 &  -1 & 0 &0  \\
  0&    0 & 0 &  0 & -1 &0  \\
  0&    0 & 0 &  0 & 0 &-1  \\
\end{array} } \right],
\end{equation}
and
\begin{equation}
\rho^c = 
\left[ {\begin{array}{cc}
\rho_{\bar\nu}  & X^*   \\
X^T  &  \rho_{\nu}     \\
\end{array} } \right].
\end{equation}

\begin{figure} [t!]
\begin{center}
\includegraphics*[width=.48\textwidth,trim=0 10 10 0,clip]{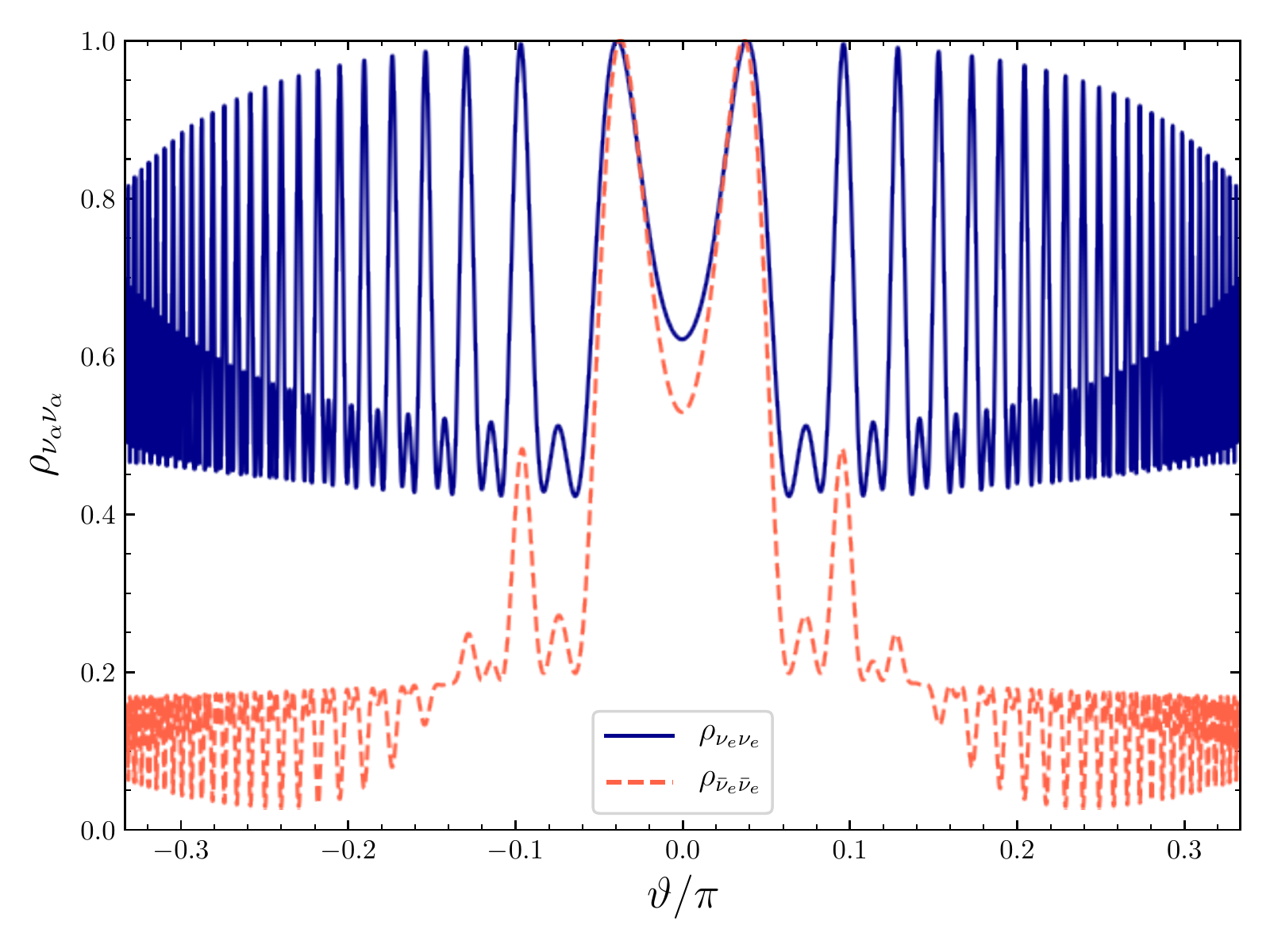}
\end{center}
\caption{  Angular distributions of the survival probabilities of
 neutrinos and antineutrinos corresponding to the calculation in 
 Fig~\ref{fig:3f} with $n_{\bar\nu_e}/n_{\nu_e}=0.7$, and at 
 $r = 1000\ \rm{cm}$.} 
\label{fig:angular}
\end{figure}

\subsubsection{Results}
Collective neutrino oscillations in strong magnetic field in the three-flavor
scenario is very similar to the one in the two-flavor scenario. In particular,
the angle-averaged survival probabilities can reach some sort of flavor 
equilibrium, as indicated in Fig.~\ref{fig:3f}. However, the magnetic
term is more complicated in the three-flavor scenario. Thus, different
flavors can, in general,  reach different  equilibrium values. For example,
in our calculations with $n_{\nu_\mu} = n_{\bar\nu_\mu} = 
n_{\nu_\tau} = n_{\bar\nu_\tau}$, we observed that
$\nu_e$ and $\bar\nu_e$ reach a flavor equilibrium in which
\begin{equation}
\begin{split}
\rho_{\nu_e \nu_e} \simeq   \frac{ n_{\nu_e}  + n_{\bar{\nu}_\mu} +  n_{\bar{\nu}_\tau} }{3}\\
\rho_{\bar{\nu}_e \bar{\nu}_e} \simeq  \frac{ n_{\bar{\nu}_e} + n_{\nu_\mu} + n_{\nu_\tau}  }{3}.
\end{split}
\end{equation}
This can be explained by noting that the magnetic term couples $\nu_e$ to $\bar\nu_\mu$,$\bar\nu_\tau$
and $\bar\nu_e$ to $\nu_\mu$,$\nu_\tau$.   
In general, the  equilibrium values of different neutrino species are only functions 
of  the neutrino number densities
but  independent of other quantities such as
the mass term in the Hamiltonian, $\vartheta_{\rm{max}}$, the
matter density and so on.

 Although individual neutrino (angle) beams can experience
large amplitude flavor oscillations, 
the rapid variations of the angular distributions of 
 neutrino  survival probabilities, as shown in Fig. \ref{fig:angular}, 
 allow neutrinos to reach a flavor equilibrium
with relatively small amplitude oscillations around the equilibrium value.

\section{Conclusion}

We have studied collective  oscillations of Majorana neutrinos in a dense 
neutrino gas in the presence of strong magnetic fields. 
Such physical environment  is thought to exist
in  NSM remnants and 
magneto-rotational CCSNe.

Collective oscillations of Majorana neutrinos can lead to
a sort of  approximate flavor equilibrium 
in the presence of strong magnetic field provided that the neutrino
 transition magnetic moment is strong enough, i.e. when 
 $\Omega = \mu_\nu B_{\rm{T}}$ is comparable to the other 
  terms in the Hamiltonian.
  The equilibrium state is 
  determined 
 by the  number densities 
of the coupled (through the magnetic term) neutrino/antineutrino species.

In the presence of  nonzero Majorana  neutrino
magnetic moment, although the total number of neutrinos plus antineutrinos is conserved,
the number of neutrinos and antineutrinos are not individually conserved.
This is different from the case of  usual  collective neutrino oscillations 
in the absence of magnetic fields (and other beyond SM terms).
We do not consider the case of Dirac neutrinos in this study
where the total number of active neutrinos can also be changed 
due to the possibility of  transition between active and sterile
neutrinos.

In our calculations, the magnetic coupling term can play a noticeable role
only if  $\Omega$ is not much smaller than the other terms in the
Hamiltonian. This is different from the observation of 
Refs.~\cite{deGouvea:2012hg, deGouvea:2013zp} where the presence of the magnetic term can significantly modify
collective neutrino oscillations even if it is many orders of magnitude
smaller than the other terms in the Hamiltonian. 
This means that one should observe a remarkable impact
from the magnetic term 
even if the length scale associated with 
$\Omega$ is orders of magnitude larger than the size of the 
SN\footnote{In the case of  usual  collective neutrino oscillations 
(in the absence of magnetic fields), the presence of the off-diagonal
term can have a switch-on effect on flavor conversion. Thus, even
an infinitesimal nonzero off-diagonal term ($\theta_{\rm{v}} \ll 1$) can significantly
affect flavor evolution of neutrinos. This arises from the flavor instabilities 
induced by neutrino-neutrino interactions. However, one should not generally expect
a similar effect from  small $\nu-\bar\nu$ transition amplitude due to the presence
of $\Omega$
unless neutrino-neutrino interaction has some contribution to $\nu-\bar\nu$ 
transition, as in Refs.~\cite{deGouvea:2012hg, deGouvea:2013zp}.}.
Note, however, that any comparison between the results presented here
and the ones in Refs.~\cite{deGouvea:2012hg, deGouvea:2013zp}
 must be made with great caution since the
employed models are different in several respects.

 Apart from providing an example of a physical situation in which  
collective neutrino oscillations can lead to generic flavor equilibrium, our
findings provide  useful insight on how the presence of strong 
 lepton number violating channels 
 can impact collective neutrino oscillations. Our results can have important 
 implications for the  physics of the most extreme astrophysical 
 environments  such as NSM remnants and 
 magneto-rotational CCSNe.



\section*{Acknowledgments}
I would like to thank  C.~Volpe, S.~Shalgar and M.~Obergaulinger for valuable
discussions and H.~Duan for insightful conversations and his helpful comments on the manuscript.
 I am also grateful to V.~Cirigliano for providing me with his notes on
neutrino quantum kinetics. This work is partially  supported by ”Physique 
fondamentale et ondes gravitationelles” (PhysFOG) of the Observatoire de Paris.


\bibliographystyle{elsarticle-num}
\bibliography{fast_modes}

\begin{thebibliography}{10}
\expandafter\ifx\csname url\endcsname\relax
  \def\url#1{\texttt{#1}}\fi
\expandafter\ifx\csname urlprefix\endcsname\relax\def\urlprefix{URL }\fi
\expandafter\ifx\csname href\endcsname\relax
  \def\href#1#2{#2} \def\path#1{#1}\fi

\bibitem{Colgate:1966ax}
S.~A. Colgate, R.~H. White, {The Hydrodynamic Behavior of Supernovae
  Explosions}, Astrophys. J. 143 (1966) 626.
\newblock \href {http://dx.doi.org/10.1086/148549} {\path{doi:10.1086/148549}}.

\bibitem{Bethe:1984ux}
H.~A. Bethe, J.~R. Wilson, {Revival of a stalled supernova shock by neutrino
  heating}, Astrophys. J. 295 (1985) 14--23.
\newblock \href {http://dx.doi.org/10.1086/163343} {\path{doi:10.1086/163343}}.

\bibitem{Janka:2012wk}
H.-T. Janka, {Explosion Mechanisms of Core-Collapse Supernovae}, Ann. Rev.
  Nucl. Part. Sci. 62 (2012) 407--451.
\newblock \href {http://arxiv.org/abs/1206.2503} {\path{arXiv:1206.2503}},
  \href {http://dx.doi.org/10.1146/annurev-nucl-102711-094901}
  {\path{doi:10.1146/annurev-nucl-102711-094901}}.

\bibitem{Burrows:2012ew}
A.~Burrows, {Colloquium: Perspectives on core-collapse supernova theory}, Rev.
  Mod. Phys. 85 (2013) 245.
\newblock \href {http://arxiv.org/abs/1210.4921} {\path{arXiv:1210.4921}},
  \href {http://dx.doi.org/10.1103/RevModPhys.85.245}
  {\path{doi:10.1103/RevModPhys.85.245}}.

\bibitem{Qian:1996xt}
Y.~Z. Qian, S.~E. Woosley, {Nucleosynthesis in neutrino driven winds: 1. The
  Physical conditions}, Astrophys. J. 471 (1996) 331--351.
\newblock \href {http://arxiv.org/abs/astro-ph/9611094}
  {\path{arXiv:astro-ph/9611094}}, \href {http://dx.doi.org/10.1086/177973}
  {\path{doi:10.1086/177973}}.

\bibitem{Pastor:2002we}
S.~Pastor, G.~Raffelt, Flavor oscillations in the supernova hot bubble region:
  Nonlinear effects of neutrino background, Phys. Rev. Lett. 89 (2002) 191101.
\newblock \href {http://arxiv.org/abs/astro-ph/0207281}
  {\path{arXiv:astro-ph/0207281}}.

\bibitem{duan:2006an}
H.~Duan, G.~M. Fuller, J.~Carlson, Y.-Z. Qian, {Simulation of Coherent
  Non-Linear Neutrino Flavor Transformation in the Supernova Environment. 1.
  Correlated Neutrino Trajectories}, Phys. Rev. D74 (2006) 105014.
\newblock \href {http://arxiv.org/abs/astro-ph/0606616}
  {\path{arXiv:astro-ph/0606616}}, \href
  {http://dx.doi.org/10.1103/PhysRevD.74.105014}
  {\path{doi:10.1103/PhysRevD.74.105014}}.

\bibitem{duan:2006jv}
H.~Duan, G.~M. Fuller, J.~Carlson, Y.-Z. Qian, {Coherent Development of
  Neutrino Flavor in the Supernova Environment}, Phys. Rev. Lett. 97 (2006)
  241101.
\newblock \href {http://arxiv.org/abs/astro-ph/0608050}
  {\path{arXiv:astro-ph/0608050}}, \href
  {http://dx.doi.org/10.1103/PhysRevLett.97.241101}
  {\path{doi:10.1103/PhysRevLett.97.241101}}.

\bibitem{duan:2010bg}
H.~Duan, G.~M. Fuller, Y.-Z. Qian, {Collective Neutrino Oscillations}, Ann.
  Rev. Nucl. Part. Sci. 60 (2010) 569--594.
\newblock \href {http://arxiv.org/abs/1001.2799} {\path{arXiv:1001.2799}},
  \href {http://dx.doi.org/10.1146/annurev.nucl.012809.104524}
  {\path{doi:10.1146/annurev.nucl.012809.104524}}.

\bibitem{Chakraborty:2016yeg}
S.~Chakraborty, R.~Hansen, I.~Izaguirre, G.~Raffelt, {Collective neutrino
  flavor conversion: Recent developments}, Nucl. Phys. B908 (2016) 366--381.
\newblock \href {http://arxiv.org/abs/1602.02766} {\path{arXiv:1602.02766}},
  \href {http://dx.doi.org/10.1016/j.nuclphysb.2016.02.012}
  {\path{doi:10.1016/j.nuclphysb.2016.02.012}}.

\bibitem{duan:2007sh}
H.~Duan, G.~M. Fuller, J.~Carlson, Y.-Z. Qian, {Flavor Evolution of the
  Neutronization Neutrino Burst from an O-Ne-Mg Core-Collapse Supernova}, Phys.
  Rev. Lett. 100 (2008) 021101.
\newblock \href {http://arxiv.org/abs/0710.1271} {\path{arXiv:0710.1271}},
  \href {http://dx.doi.org/10.1103/PhysRevLett.100.021101}
  {\path{doi:10.1103/PhysRevLett.100.021101}}.

\bibitem{dasgupta:2009mg}
B.~Dasgupta, A.~Dighe, G.~G. Raffelt, A.~Y. Smirnov, {Multiple Spectral Splits
  of Supernova Neutrinos}, Phys. Rev. Lett. 103 (2009) 051105.
\newblock \href {http://arxiv.org/abs/0904.3542} {\path{arXiv:0904.3542}},
  \href {http://dx.doi.org/10.1103/PhysRevLett.103.051105}
  {\path{doi:10.1103/PhysRevLett.103.051105}}.

\bibitem{Galais:2011gh}
S.~Galais, C.~Volpe, {The neutrino spectral split in core-collapse supernovae:
  a magnetic resonance phenomenon}, Phys. Rev. D84 (2011) 085005.
\newblock \href {http://arxiv.org/abs/1103.5302} {\path{arXiv:1103.5302}},
  \href {http://dx.doi.org/10.1103/PhysRevD.84.085005}
  {\path{doi:10.1103/PhysRevD.84.085005}}.

\bibitem{Duan:2007bt}
H.~Duan, G.~M. Fuller, J.~Carlson, Y.-Z. Qian, {Neutrino Mass Hierarchy and
  Stepwise Spectral Swapping of Supernova Neutrino Flavors}, Phys. Rev. Lett.
  99 (2007) 241802.
\newblock \href {http://arxiv.org/abs/0707.0290} {\path{arXiv:0707.0290}},
  \href {http://dx.doi.org/10.1103/PhysRevLett.99.241802}
  {\path{doi:10.1103/PhysRevLett.99.241802}}.

\bibitem{Duan:2015cqa}
H.~Duan, {Collective neutrino oscillations and spontaneous symmetry breaking},
  Int. J. Mod. Phys. E24~(09) (2015) 1541008.
\newblock \href {http://arxiv.org/abs/1506.08629} {\path{arXiv:1506.08629}},
  \href {http://dx.doi.org/10.1142/S0218301315410086}
  {\path{doi:10.1142/S0218301315410086}}.

\bibitem{Malkus:2012ts}
A.~Malkus, J.~P. Kneller, G.~C. McLaughlin, R.~Surman, {Neutrino oscillations
  above black hole accretion disks: disks with electron-flavor emission}, Phys.
  Rev. D86 (2012) 085015.
\newblock \href {http://arxiv.org/abs/1207.6648} {\path{arXiv:1207.6648}},
  \href {http://dx.doi.org/10.1103/PhysRevD.86.085015}
  {\path{doi:10.1103/PhysRevD.86.085015}}.

\bibitem{Malkus:2014iqa}
A.~Malkus, A.~Friedland, G.~C. McLaughlin, {Matter-Neutrino Resonance Above
  Merging Compact Objects}\href {http://arxiv.org/abs/1403.5797}
  {\path{arXiv:1403.5797}}.

\bibitem{Malkus:2015mda}
A.~Malkus, G.~C. McLaughlin, R.~Surman, {Symmetric and Standard Matter-Neutrino
  Resonances Above Merging Compact Objects}, Phys. Rev. D93~(4) (2016) 045021.
\newblock \href {http://arxiv.org/abs/1507.00946} {\path{arXiv:1507.00946}},
  \href {http://dx.doi.org/10.1103/PhysRevD.93.045021}
  {\path{doi:10.1103/PhysRevD.93.045021}}.

\bibitem{Vlasenko:2018irq}
A.~Vlasenko, G.~C. McLaughlin, {Matter-neutrino resonance in a multiangle
  neutrino bulb model}, Phys. Rev. D97~(8) (2018) 083011.
\newblock \href {http://arxiv.org/abs/1801.07813} {\path{arXiv:1801.07813}},
  \href {http://dx.doi.org/10.1103/PhysRevD.97.083011}
  {\path{doi:10.1103/PhysRevD.97.083011}}.

\bibitem{Shalgar:2017pzd}
S.~Shalgar, {Multi-angle calculation of the matter-neutrino resonance near an
  accretion disk}, JCAP 1802~(02) (2018) 010.
\newblock \href {http://arxiv.org/abs/1707.07692} {\path{arXiv:1707.07692}},
  \href {http://dx.doi.org/10.1088/1475-7516/2018/02/010}
  {\path{doi:10.1088/1475-7516/2018/02/010}}.

\bibitem{Tian:2017xbr}
J.~Y. Tian, A.~V. Patwardhan, G.~M. Fuller, {Neutrino Flavor Evolution in
  Neutron Star Mergers}, Phys. Rev. D96~(4) (2017) 043001.
\newblock \href {http://arxiv.org/abs/1703.03039} {\path{arXiv:1703.03039}},
  \href {http://dx.doi.org/10.1103/PhysRevD.96.043001}
  {\path{doi:10.1103/PhysRevD.96.043001}}.

\bibitem{Frensel:2016fge}
M.~Frensel, M.-R. Wu, C.~Volpe, A.~Perego, {Neutrino Flavor Evolution in Binary
  Neutron Star Merger Remnants}, Phys. Rev. D95~(2) (2017) 023011.
\newblock \href {http://arxiv.org/abs/1607.05938} {\path{arXiv:1607.05938}},
  \href {http://dx.doi.org/10.1103/PhysRevD.95.023011}
  {\path{doi:10.1103/PhysRevD.95.023011}}.

\bibitem{Zhu:2016mwa}
Y.-L. Zhu, A.~Perego, G.~C. McLaughlin, {Matter Neutrino Resonance Transitions
  above a Neutron Star Merger Remnant}, Phys. Rev. D94~(10) (2016) 105006.
\newblock \href {http://arxiv.org/abs/1607.04671} {\path{arXiv:1607.04671}},
  \href {http://dx.doi.org/10.1103/PhysRevD.94.105006}
  {\path{doi:10.1103/PhysRevD.94.105006}}.

\bibitem{Wu:2015fga}
M.-R. Wu, H.~Duan, Y.-Z. Qian, {Physics of neutrino flavor transformation
  through matter–neutrino resonances}, Phys. Lett. B752 (2016) 89--94.
\newblock \href {http://arxiv.org/abs/1509.08975} {\path{arXiv:1509.08975}},
  \href {http://dx.doi.org/10.1016/j.physletb.2015.11.027}
  {\path{doi:10.1016/j.physletb.2015.11.027}}.

\bibitem{Chatelain:2016xva}
A.~Chatelain, C.~Volpe, {Helicity coherence in binary neutron star mergers and
  non-linear feedback}, Phys. Rev. D95~(4) (2017) 043005.
\newblock \href {http://arxiv.org/abs/1611.01862} {\path{arXiv:1611.01862}},
  \href {http://dx.doi.org/10.1103/PhysRevD.95.043005}
  {\path{doi:10.1103/PhysRevD.95.043005}}.

\bibitem{Chatelain:2017yxx}
A.~Chatelain, M.~C. Volpe, {Neutrino propagation in binary neutron star mergers
  in presence of nonstandard interactions}, Phys. Rev. D97~(2) (2018) 023014.
\newblock \href {http://arxiv.org/abs/1710.11518} {\path{arXiv:1710.11518}},
  \href {http://dx.doi.org/10.1103/PhysRevD.97.023014}
  {\path{doi:10.1103/PhysRevD.97.023014}}.

\bibitem{DelfanAzari:2019tez}
M.~Delfan~Azari, S.~Yamada, T.~Morinaga, H.~Nagakura, S.~Furusawa, A.~Harada,
  H.~Okawa, W.~Iwakami, K.~Sumiyoshi, {Fast collective neutrino oscillations
  inside the neutrino sphere in core-collapse supernovae}\href
  {http://arxiv.org/abs/1910.06176} {\path{arXiv:1910.06176}}.

\bibitem{Abbar:2019zoq}
S.~Abbar, H.~Duan, K.~Sumiyoshi, T.~Takiwaki, M.~C. Volpe, {Fast Neutrino
  Flavor Conversion Modes in Multidimensional Core-collapse Supernova Models:
  the Role of the Asymmetric Neutrino Distributions}\href
  {http://arxiv.org/abs/1911.01983} {\path{arXiv:1911.01983}}.

\bibitem{Glas:2019ijo}
R.~Glas, H.~T. Janka, F.~Capozzi, M.~Sen, B.~Dasgupta, A.~Mirizzi, G.~Sigl,
  {Fast Neutrino Flavor Instability in the Neutron-star Convection Layer of
  Three-dimensional Supernova Models}\href {http://arxiv.org/abs/1912.00274}
  {\path{arXiv:1912.00274}}.

\bibitem{raffelt:2013rqa}
G.~Raffelt, S.~Sarikas, D.~d.~S. Seixas, {Axial symmetry breaking in
  self-induced flavor conversion of supernova neutrino fluxes}, Phys. Rev.
  Lett. 111 (2013) 091101.
\newblock \href {http://arxiv.org/abs/1305.7140} {\path{arXiv:1305.7140}},
  \href {http://dx.doi.org/10.1103/PhysRevLett.111.091101}
  {\path{doi:10.1103/PhysRevLett.111.091101}}.

\bibitem{duan:2013kba}
H.~Duan, {Flavor Oscillation Modes In Dense Neutrino Media}, Phys. Rev. D88
  (2013) 125008.
\newblock \href {http://arxiv.org/abs/1309.7377} {\path{arXiv:1309.7377}},
  \href {http://dx.doi.org/10.1103/PhysRevD.88.125008}
  {\path{doi:10.1103/PhysRevD.88.125008}}.

\bibitem{duan:2014gfa}
H.~Duan, S.~Shalgar, {Flavor instabilities in the neutrino line model}, Phys.
  Lett. B747 (2015) 139--143.
\newblock \href {http://arxiv.org/abs/1412.7097} {\path{arXiv:1412.7097}},
  \href {http://dx.doi.org/10.1016/j.physletb.2015.05.057}
  {\path{doi:10.1016/j.physletb.2015.05.057}}.

\bibitem{abbar:2015mca}
S.~Abbar, H.~Duan, S.~Shalgar, {Flavor instabilities in the multiangle neutrino
  line model}, Phys. Rev. D92~(6) (2015) 065019.
\newblock \href {http://arxiv.org/abs/1507.08992} {\path{arXiv:1507.08992}},
  \href {http://dx.doi.org/10.1103/PhysRevD.92.065019}
  {\path{doi:10.1103/PhysRevD.92.065019}}.

\bibitem{Abbar:2015fwa}
S.~Abbar, H.~Duan, {Neutrino flavor instabilities in a time-dependent supernova
  model}, Phys. Lett. B751 (2015) 43--47.
\newblock \href {http://arxiv.org/abs/1509.01538} {\path{arXiv:1509.01538}},
  \href {http://dx.doi.org/10.1016/j.physletb.2015.10.019}
  {\path{doi:10.1016/j.physletb.2015.10.019}}.

\bibitem{chakraborty:2015tfa}
S.~Chakraborty, R.~S. Hansen, I.~Izaguirre, G.~Raffelt, {Self-induced flavor
  conversion of supernova neutrinos on small scales}, JCAP 1601~(01) (2016)
  028.
\newblock \href {http://arxiv.org/abs/1507.07569} {\path{arXiv:1507.07569}},
  \href {http://dx.doi.org/10.1088/1475-7516/2016/01/028}
  {\path{doi:10.1088/1475-7516/2016/01/028}}.

\bibitem{Dasgupta:2015iia}
B.~Dasgupta, A.~Mirizzi, {Temporal Instability Enables Neutrino Flavor
  Conversions Deep Inside Supernovae}, Phys. Rev. D92~(12) (2015) 125030.
\newblock \href {http://arxiv.org/abs/1509.03171} {\path{arXiv:1509.03171}},
  \href {http://dx.doi.org/10.1103/PhysRevD.92.125030}
  {\path{doi:10.1103/PhysRevD.92.125030}}.

\bibitem{Mirizzi:2015fva}
A.~Mirizzi, G.~Mangano, N.~Saviano, {Self-induced flavor instabilities of a
  dense neutrino stream in a two-dimensional model}, Phys. Rev. D92~(2) (2015)
  021702.
\newblock \href {http://arxiv.org/abs/1503.03485} {\path{arXiv:1503.03485}},
  \href {http://dx.doi.org/10.1103/PhysRevD.92.021702}
  {\path{doi:10.1103/PhysRevD.92.021702}}.

\bibitem{Martin:2019gxb}
J.~D. Martin, C.~Yi, H.~Duan, {Dynamic fast flavor oscillation waves in dense
  neutrino gases}, Phys. Lett. B800 (2020) 135088.
\newblock \href {http://arxiv.org/abs/1909.05225} {\path{arXiv:1909.05225}},
  \href {http://dx.doi.org/10.1016/j.physletb.2019.135088}
  {\path{doi:10.1016/j.physletb.2019.135088}}.

\bibitem{Martin:2019kgi}
J.~D. Martin, S.~Abbar, H.~Duan, {Nonlinear flavor development of a
  two-dimensional neutrino gas}, Phys. Rev. D100~(2) (2019) 023016.
\newblock \href {http://arxiv.org/abs/1904.08877} {\path{arXiv:1904.08877}},
  \href {http://dx.doi.org/10.1103/PhysRevD.100.023016}
  {\path{doi:10.1103/PhysRevD.100.023016}}.

\bibitem{Sawyer:2005jk}
R.~F. Sawyer, {Speed-up of neutrino transformations in a supernova
  environment}, Phys. Rev. D72 (2005) 045003.
\newblock \href {http://arxiv.org/abs/hep-ph/0503013}
  {\path{arXiv:hep-ph/0503013}}, \href
  {http://dx.doi.org/10.1103/PhysRevD.72.045003}
  {\path{doi:10.1103/PhysRevD.72.045003}}.

\bibitem{Sawyer:2015dsa}
R.~F. Sawyer, {Neutrino cloud instabilities just above the neutrino sphere of a
  supernova}, Phys. Rev. Lett. 116~(8) (2016) 081101.
\newblock \href {http://arxiv.org/abs/1509.03323} {\path{arXiv:1509.03323}},
  \href {http://dx.doi.org/10.1103/PhysRevLett.116.081101}
  {\path{doi:10.1103/PhysRevLett.116.081101}}.

\bibitem{Chakraborty:2016lct}
S.~Chakraborty, R.~S. Hansen, I.~Izaguirre, G.~Raffelt, {Self-induced neutrino
  flavor conversion without flavor mixing}, JCAP 1603~(03) (2016) 042.
\newblock \href {http://arxiv.org/abs/1602.00698} {\path{arXiv:1602.00698}},
  \href {http://dx.doi.org/10.1088/1475-7516/2016/03/042}
  {\path{doi:10.1088/1475-7516/2016/03/042}}.

\bibitem{Izaguirre:2016gsx}
I.~Izaguirre, G.~Raffelt, I.~Tamborra, {Fast Pairwise Conversion of Supernova
  Neutrinos: A Dispersion-Relation Approach}, Phys. Rev. Lett. 118~(2) (2017)
  021101.
\newblock \href {http://arxiv.org/abs/1610.01612} {\path{arXiv:1610.01612}},
  \href {http://dx.doi.org/10.1103/PhysRevLett.118.021101}
  {\path{doi:10.1103/PhysRevLett.118.021101}}.

\bibitem{Wu:2017qpc}
M.-R. Wu, I.~Tamborra, {Fast neutrino conversions: Ubiquitous in compact binary
  merger remnants}, Phys. Rev. D95~(10) (2017) 103007.
\newblock \href {http://arxiv.org/abs/1701.06580} {\path{arXiv:1701.06580}},
  \href {http://dx.doi.org/10.1103/PhysRevD.95.103007}
  {\path{doi:10.1103/PhysRevD.95.103007}}.

\bibitem{Capozzi:2017gqd}
F.~Capozzi, B.~Dasgupta, E.~Lisi, A.~Marrone, A.~Mirizzi, {Fast flavor
  conversions of supernova neutrinos: Classifying instabilities via dispersion
  relations}, Phys. Rev. D96~(4) (2017) 043016.
\newblock \href {http://arxiv.org/abs/1706.03360} {\path{arXiv:1706.03360}},
  \href {http://dx.doi.org/10.1103/PhysRevD.96.043016}
  {\path{doi:10.1103/PhysRevD.96.043016}}.

\bibitem{Richers:2019grc}
S.~A. Richers, G.~C. McLaughlin, J.~P. Kneller, A.~Vlasenko, {Neutrino Quantum
  Kinetics in Compact Objects}, Phys. Rev. D99~(12) (2019) 123014.
\newblock \href {http://arxiv.org/abs/1903.00022} {\path{arXiv:1903.00022}},
  \href {http://dx.doi.org/10.1103/PhysRevD.99.123014}
  {\path{doi:10.1103/PhysRevD.99.123014}}.

\bibitem{Dasgupta:2016dbv}
B.~Dasgupta, A.~Mirizzi, M.~Sen, {Fast neutrino flavor conversions near the
  supernova core with realistic flavor-dependent angular distributions}, JCAP
  1702~(02) (2017) 019.
\newblock \href {http://arxiv.org/abs/1609.00528} {\path{arXiv:1609.00528}},
  \href {http://dx.doi.org/10.1088/1475-7516/2017/02/019}
  {\path{doi:10.1088/1475-7516/2017/02/019}}.

\bibitem{Abbar:2017pkh}
S.~Abbar, H.~Duan, {Fast neutrino flavor conversion: roles of dense matter and
  spectrum crossing}, Phys. Rev. D98~(4) (2018) 043014.
\newblock \href {http://arxiv.org/abs/1712.07013} {\path{arXiv:1712.07013}},
  \href {http://dx.doi.org/10.1103/PhysRevD.98.043014}
  {\path{doi:10.1103/PhysRevD.98.043014}}.

\bibitem{Abbar:2018beu}
S.~Abbar, M.~C. Volpe, {On Fast Neutrino Flavor Conversion Modes in the
  Nonlinear Regime}, Phys. Lett. B790 (2019) 545--550.
\newblock \href {http://arxiv.org/abs/1811.04215} {\path{arXiv:1811.04215}},
  \href {http://dx.doi.org/10.1016/j.physletb.2019.02.002}
  {\path{doi:10.1016/j.physletb.2019.02.002}}.

\bibitem{Capozzi:2018clo}
F.~Capozzi, B.~Dasgupta, A.~Mirizzi, M.~Sen, G.~Sigl, {Collisional triggering
  of fast flavor conversions of supernova neutrinos}, Phys. Rev. Lett. 122~(9)
  (2019) 091101.
\newblock \href {http://arxiv.org/abs/1808.06618} {\path{arXiv:1808.06618}},
  \href {http://dx.doi.org/10.1103/PhysRevLett.122.091101}
  {\path{doi:10.1103/PhysRevLett.122.091101}}.

\bibitem{Capozzi:2019lso}
F.~Capozzi, G.~Raffelt, T.~Stirner, {Fast Neutrino Flavor Conversion:
  Collective Motion vs. Decoherence}, JCAP 1909 (2019) 002.
\newblock \href {http://arxiv.org/abs/1906.08794} {\path{arXiv:1906.08794}},
  \href {http://dx.doi.org/10.1088/1475-7516/2019/09/002}
  {\path{doi:10.1088/1475-7516/2019/09/002}}.

\bibitem{Doring:2019axc}
C.~Döring, R.~S.~L. Hansen, M.~Lindner, {Stability of three neutrino flavor
  conversion in supernovae}, JCAP 1908 (2019) 003.
\newblock \href {http://arxiv.org/abs/1905.03647} {\path{arXiv:1905.03647}},
  \href {http://dx.doi.org/10.1088/1475-7516/2019/08/003}
  {\path{doi:10.1088/1475-7516/2019/08/003}}.

\bibitem{Chakraborty:2019wxe}
M.~Chakraborty, S.~Chakraborty, {Three flavor neutrino conversions in
  supernovae: Slow $\&$ Fast instabilities}\href
  {http://arxiv.org/abs/1909.10420} {\path{arXiv:1909.10420}}.

\bibitem{Johns:2019izj}
L.~Johns, H.~Nagakura, G.~M. Fuller, A.~Burrows, {Neutrino oscillations in
  supernovae: angular moments and fast instabilities}\href
  {http://arxiv.org/abs/1910.05682} {\path{arXiv:1910.05682}}.

\bibitem{Shalgar:2019qwg}
S.~Shalgar, I.~Padilla-Gay, I.~Tamborra, {Neutrino propagation hinders fast
  pairwise flavor conversions}\href {http://arxiv.org/abs/1911.09110}
  {\path{arXiv:1911.09110}}.

\bibitem{Cherry:2019vkv}
J.~F. Cherry, G.~M. Fuller, S.~Horiuchi, K.~Kotake, T.~Takiwaki, T.~Fischer,
  {Time of Flight and Supernova Progenitor Effects on the Neutrino Halo}\href
  {http://arxiv.org/abs/1912.11489} {\path{arXiv:1912.11489}}.

\bibitem{Giunti:2008ve}
C.~Giunti, A.~Studenikin, {Neutrino electromagnetic properties}, Phys. Atom.
  Nucl. 72 (2009) 2089--2125.
\newblock \href {http://arxiv.org/abs/0812.3646} {\path{arXiv:0812.3646}},
  \href {http://dx.doi.org/10.1134/S1063778809120126}
  {\path{doi:10.1134/S1063778809120126}}.

\bibitem{Broggini:2012df}
C.~Broggini, C.~Giunti, A.~Studenikin, {Electromagnetic Properties of
  Neutrinos}, Adv. High Energy Phys. 2012 (2012) 459526.
\newblock \href {http://arxiv.org/abs/1207.3980} {\path{arXiv:1207.3980}},
  \href {http://dx.doi.org/10.1155/2012/459526}
  {\path{doi:10.1155/2012/459526}}.

\bibitem{Studenikin:2016ykv}
A.~Studenikin, {Status and perspectives of neutrino magnetic moments}, J. Phys.
  Conf. Ser. 718~(6) (2016) 062076.
\newblock \href {http://arxiv.org/abs/1603.00337} {\path{arXiv:1603.00337}},
  \href {http://dx.doi.org/10.1088/1742-6596/718/6/062076}
  {\path{doi:10.1088/1742-6596/718/6/062076}}.

\bibitem{Mosta:2015ucs}
P.~Mösta, C.~D. Ott, D.~Radice, L.~F. Roberts, E.~Schnetter, R.~Haas, {A large
  scale dynamo and magnetoturbulence in rapidly rotating core-collapse
  supernovae}, Nature 528 (2015) 376.
\newblock \href {http://arxiv.org/abs/1512.00838} {\path{arXiv:1512.00838}},
  \href {http://dx.doi.org/10.1038/nature15755}
  {\path{doi:10.1038/nature15755}}.

\bibitem{Fujikawa:1980yx}
K.~Fujikawa, R.~Shrock, {The Magnetic Moment of a Massive Neutrino and Neutrino
  Spin Rotation}, Phys. Rev. Lett. 45 (1980) 963.
\newblock \href {http://dx.doi.org/10.1103/PhysRevLett.45.963}
  {\path{doi:10.1103/PhysRevLett.45.963}}.

\bibitem{Lindner:2017uvt}
M.~Lindner, B.~Radovčić, J.~Welter, {Revisiting Large Neutrino Magnetic
  Moments}, JHEP 07 (2017) 139.
\newblock \href {http://arxiv.org/abs/1706.02555} {\path{arXiv:1706.02555}},
  \href {http://dx.doi.org/10.1007/JHEP07(2017)139}
  {\path{doi:10.1007/JHEP07(2017)139}}.

\bibitem{Bell:2005kz}
N.~F. Bell, V.~Cirigliano, M.~J. Ramsey-Musolf, P.~Vogel, M.~B. Wise, {How
  magnetic is the Dirac neutrino?}, Phys. Rev. Lett. 95 (2005) 151802.
\newblock \href {http://arxiv.org/abs/hep-ph/0504134}
  {\path{arXiv:hep-ph/0504134}}, \href
  {http://dx.doi.org/10.1103/PhysRevLett.95.151802}
  {\path{doi:10.1103/PhysRevLett.95.151802}}.

\bibitem{Davidson:2005cs}
S.~Davidson, M.~Gorbahn, A.~Santamaria, {From transition magnetic moments to
  majorana neutrino masses}, Phys. Lett. B626 (2005) 151--160.
\newblock \href {http://arxiv.org/abs/hep-ph/0506085}
  {\path{arXiv:hep-ph/0506085}}, \href
  {http://dx.doi.org/10.1016/j.physletb.2005.08.086}
  {\path{doi:10.1016/j.physletb.2005.08.086}}.

\bibitem{Canas:2015yoa}
B.~C. Canas, O.~G. Miranda, A.~Parada, M.~Tortola, J.~W.~F. Valle, {Updating
  neutrino magnetic moment constraints}, Phys. Lett. B753 (2016) 191--198,
  [Addendum: Phys. Lett.B757,568(2016)].
\newblock \href {http://arxiv.org/abs/1510.01684} {\path{arXiv:1510.01684}},
  \href {http://dx.doi.org/10.1016/j.physletb.2016.03.078,
  10.1016/j.physletb.2015.12.011} {\path{doi:10.1016/j.physletb.2016.03.078,
  10.1016/j.physletb.2015.12.011}}.

\bibitem{Beda:2012zz}
A.~G. Beda, V.~B. Brudanin, V.~G. Egorov, D.~V. Medvedev, V.~S. Pogosov, M.~V.
  Shirchenko, A.~S. Starostin, {The results of search for the neutrino magnetic
  moment in GEMMA experiment}, Adv. High Energy Phys. 2012 (2012) 350150.
\newblock \href {http://dx.doi.org/10.1155/2012/350150}
  {\path{doi:10.1155/2012/350150}}.

\bibitem{Kolb:1981mc}
E.~W. Kolb, D.~L. Tubbs, D.~A. Dicus, {Lepton Number Violation, Majorana
  Neutrinos, and Supernovae}, Astrophys. J. 255 (1982) L57.
\newblock \href {http://dx.doi.org/10.1086/183769} {\path{doi:10.1086/183769}}.

\bibitem{Schechter:1981hw}
J.~Schechter, J.~W.~F. Valle, {Majorana Neutrinos and Magnetic Fields}, Phys.
  Rev. D24 (1981) 1883--1889, [Erratum: Phys. Rev.D25,283(1982)].
\newblock \href {http://dx.doi.org/10.1103/PhysRevD.25.283,
  10.1103/PhysRevD.24.1883} {\path{doi:10.1103/PhysRevD.25.283,
  10.1103/PhysRevD.24.1883}}.

\bibitem{Lim:1987tk}
C.-S. Lim, W.~J. Marciano, {Resonant Spin - Flavor Precession of Solar and
  Supernova Neutrinos}, Phys. Rev. D37 (1988) 1368--1373, [,351(1987)].
\newblock \href {http://dx.doi.org/10.1103/PhysRevD.37.1368}
  {\path{doi:10.1103/PhysRevD.37.1368}}.

\bibitem{Balantekin:1990jg}
A.~B. Balantekin, P.~J. Hatchell, F.~Loreti, {Matter Enhanced Spin Flavor
  Precession of Solar Neutrinos With Transition Magnetic Moments}, Phys. Rev.
  D41 (1990) 3583.
\newblock \href {http://dx.doi.org/10.1103/PhysRevD.41.3583}
  {\path{doi:10.1103/PhysRevD.41.3583}}.

\bibitem{Akhmedov:1992ea}
E.~K. Akhmedov, Z.~G. Berezhiani, {Implications of Majorana neutrino transition
  magnetic moments for neutrino signals from supernovae}, Nucl. Phys. B373
  (1992) 479--497.
\newblock \href {http://dx.doi.org/10.1016/0550-3213(92)90441-D}
  {\path{doi:10.1016/0550-3213(92)90441-D}}.

\bibitem{Akhmedov:1997qb}
E.~K. Akhmedov, A.~Lanza, D.~W. Sciama, {Resonant spin flavor precession of
  neutrinos and pulsar velocities}, Phys. Rev. D56 (1997) 6117--6124.
\newblock \href {http://arxiv.org/abs/hep-ph/9702436}
  {\path{arXiv:hep-ph/9702436}}, \href
  {http://dx.doi.org/10.1103/PhysRevD.56.6117}
  {\path{doi:10.1103/PhysRevD.56.6117}}.

\bibitem{Ando:2002sk}
S.~Ando, K.~Sato, {Three generation study of neutrino spin flavor conversion in
  supernova and implication for neutrino magnetic moment}, Phys. Rev. D67
  (2003) 023004.
\newblock \href {http://arxiv.org/abs/hep-ph/0211053}
  {\path{arXiv:hep-ph/0211053}}, \href
  {http://dx.doi.org/10.1103/PhysRevD.67.023004}
  {\path{doi:10.1103/PhysRevD.67.023004}}.

\bibitem{Lychkovskiy:2009pm}
O.~Lychkovskiy, S.~Blinnikov, {Spin flip of neutrinos with magnetic moment in
  core-collapse supernova}, Phys. Atom. Nucl. 73 (2010) 614--624.
\newblock \href {http://arxiv.org/abs/0905.3658} {\path{arXiv:0905.3658}},
  \href {http://dx.doi.org/10.1134/S106377881004006X}
  {\path{doi:10.1134/S106377881004006X}}.

\bibitem{Balantekin:2007xq}
A.~B. Balantekin, C.~Volpe, J.~Welzel, {Impact of the Neutrino Magnetic Moment
  on Supernova r-process Nucleosynthesis}, JCAP 0709 (2007) 016.
\newblock \href {http://arxiv.org/abs/0706.3023} {\path{arXiv:0706.3023}},
  \href {http://dx.doi.org/10.1088/1475-7516/2007/09/016}
  {\path{doi:10.1088/1475-7516/2007/09/016}}.

\bibitem{Kuznetsov:2009we}
A.~V. Kuznetsov, N.~V. Mikheev, A.~A. Okrugin, {Dirac-Neutrino Magnetic Moment
  and the Dynamics of a Supernova Explosion}, JETP Lett. 89 (2009) 97--101.
\newblock \href {http://arxiv.org/abs/0903.2321} {\path{arXiv:0903.2321}},
  \href {http://dx.doi.org/10.1134/S0021364009030011}
  {\path{doi:10.1134/S0021364009030011}}.

\bibitem{Akhmedov:2003fu}
E.~K. Akhmedov, T.~Fukuyama, {Supernova prompt neutronization neutrinos and
  neutrino magnetic moments}, JCAP 0312 (2003) 007.
\newblock \href {http://arxiv.org/abs/hep-ph/0310119}
  {\path{arXiv:hep-ph/0310119}}, \href
  {http://dx.doi.org/10.1088/1475-7516/2003/12/007}
  {\path{doi:10.1088/1475-7516/2003/12/007}}.

\bibitem{deGouvea:2012hg}
A.~de~Gouvea, S.~Shalgar, {Effect of Transition Magnetic Moments on Collective
  Supernova Neutrino Oscillations}, JCAP 1210 (2012) 027.
\newblock \href {http://arxiv.org/abs/1207.0516} {\path{arXiv:1207.0516}},
  \href {http://dx.doi.org/10.1088/1475-7516/2012/10/027}
  {\path{doi:10.1088/1475-7516/2012/10/027}}.

\bibitem{deGouvea:2013zp}
A.~de~Gouvea, S.~Shalgar, {Transition Magnetic Moments and Collective Neutrino
  Oscillations:Three-Flavor Effects and Detectability}, JCAP 1304 (2013) 018.
\newblock \href {http://arxiv.org/abs/1301.5637} {\path{arXiv:1301.5637}},
  \href {http://dx.doi.org/10.1088/1475-7516/2013/04/018}
  {\path{doi:10.1088/1475-7516/2013/04/018}}.

\bibitem{Dvornikov:2011dv}
M.~Dvornikov, {Evolution of a dense neutrino gas in matter and electromagnetic
  field}, Nucl. Phys. B855 (2012) 760--773.
\newblock \href {http://arxiv.org/abs/1108.5043} {\path{arXiv:1108.5043}},
  \href {http://dx.doi.org/10.1016/j.nuclphysb.2011.10.025}
  {\path{doi:10.1016/j.nuclphysb.2011.10.025}}.

\bibitem{Dobrynina:2016rwy}
A.~Dobrynina, A.~Kartavtsev, G.~Raffelt, {Helicity oscillations of Dirac and
  Majorana neutrinos}, Phys. Rev. D93~(12) (2016) 125030.
\newblock \href {http://arxiv.org/abs/1605.04512} {\path{arXiv:1605.04512}},
  \href {http://dx.doi.org/10.1103/PhysRevD.93.125030}
  {\path{doi:10.1103/PhysRevD.93.125030}}.

\bibitem{Kurashvili:2017zab}
P.~Kurashvili, K.~A. Kouzakov, L.~Chotorlishvili, A.~I. Studenikin,
  {Spin-flavor oscillations of ultrahigh-energy cosmic neutrinos in
  interstellar space: The role of neutrino magnetic moments}, Phys. Rev.
  D96~(10) (2017) 103017.
\newblock \href {http://arxiv.org/abs/1711.04303} {\path{arXiv:1711.04303}},
  \href {http://dx.doi.org/10.1103/PhysRevD.96.103017}
  {\path{doi:10.1103/PhysRevD.96.103017}}.

\bibitem{Sigl:1992fn}
G.~Sigl, G.~Raffelt, {General kinetic description of relativistic mixed
  neutrinos}, Nucl. Phys. B406 (1993) 423--451.
\newblock \href {http://dx.doi.org/10.1016/0550-3213(93)90175-O}
  {\path{doi:10.1016/0550-3213(93)90175-O}}.

\bibitem{Strack:2005ux}
P.~Strack, A.~Burrows, A generalized boltzmann formalism for oscillating
  neutrinos, Phys. Rev. D71 (2005) 093004.
\newblock \href {http://arxiv.org/abs/hep-ph/0504035}
  {\path{arXiv:hep-ph/0504035}}.

\bibitem{Cardall:2007zw}
C.~Y. Cardall, {Liouville equations for neutrino distribution matrices}, Phys.
  Rev. D78 (2008) 085017.
\newblock \href {http://arxiv.org/abs/0712.1188} {\path{arXiv:0712.1188}},
  \href {http://dx.doi.org/10.1103/PhysRevD.78.085017}
  {\path{doi:10.1103/PhysRevD.78.085017}}.

\bibitem{Volpe:2013jgr}
C.~Volpe, D.~Väänänen, C.~Espinoza, {Extended evolution equations for
  neutrino propagation in astrophysical and cosmological environments}, Phys.
  Rev. D87~(11) (2013) 113010.
\newblock \href {http://arxiv.org/abs/1302.2374} {\path{arXiv:1302.2374}},
  \href {http://dx.doi.org/10.1103/PhysRevD.87.113010}
  {\path{doi:10.1103/PhysRevD.87.113010}}.

\bibitem{Vlasenko:2013fja}
A.~Vlasenko, G.~M. Fuller, V.~Cirigliano, {Neutrino Quantum Kinetics}, Phys.
  Rev. D89~(10) (2014) 105004.
\newblock \href {http://arxiv.org/abs/1309.2628} {\path{arXiv:1309.2628}},
  \href {http://dx.doi.org/10.1103/PhysRevD.89.105004}
  {\path{doi:10.1103/PhysRevD.89.105004}}.

\bibitem{Cirigliano:2014aoa}
V.~Cirigliano, G.~M. Fuller, A.~Vlasenko, {A New Spin on Neutrino Quantum
  Kinetics}, Phys. Lett. B747 (2015) 27--35.
\newblock \href {http://arxiv.org/abs/1406.5558} {\path{arXiv:1406.5558}},
  \href {http://dx.doi.org/10.1016/j.physletb.2015.04.066}
  {\path{doi:10.1016/j.physletb.2015.04.066}}.

\bibitem{Vaananen:2013qja}
D.~Väänänen, C.~Volpe, {Linearizing neutrino evolution equations including
  neutrino-antineutrino pairing correlations}, Phys. Rev. D88 (2013) 065003.
\newblock \href {http://arxiv.org/abs/1306.6372} {\path{arXiv:1306.6372}},
  \href {http://dx.doi.org/10.1103/PhysRevD.88.065003}
  {\path{doi:10.1103/PhysRevD.88.065003}}.

\bibitem{Volpe:2015rla}
C.~Volpe, {Neutrino Quantum Kinetic Equations}, Int. J. Mod. Phys. E24~(09)
  (2015) 1541009.
\newblock \href {http://arxiv.org/abs/1506.06222} {\path{arXiv:1506.06222}},
  \href {http://dx.doi.org/10.1142/S0218301315410098}
  {\path{doi:10.1142/S0218301315410098}}.

\bibitem{Tanabashi:2018oca}
M.~Tanabashi, et~al., {Review of Particle Physics}, Phys. Rev. D98~(3) (2018)
  030001.
\newblock \href {http://dx.doi.org/10.1103/PhysRevD.98.030001}
  {\path{doi:10.1103/PhysRevD.98.030001}}.

\end{thebibliography}

\clearpage
\appendix

\end{document}